\documentstyle[12pt,epsf]{article}

\newcommand{\lmm}{\ln\frac{\mu^2}{m^2}}

\begin{document}

\begin{titlepage}
\noindent
%
%
\hfill TTP95--41\footnote{The complete postscript file of this
preprint, including figures, is available via anonymous ftp at
ttpux2.physik.uni-karlsruhe.de (129.13.102.139) as
/ttp95-41/ttp95-41.ps or via www at
http://www-ttp.physik.uni-karlsruhe.de/cgi-bin/preprints/
Report-no: TTP95-41.}\\
\mbox{}
\hfill hep-ph/9511430\\
\mbox{}
\hfill November 1995\\

\vspace{.5cm}

\begin{center}
\begin{LARGE}
 {\bf Heavy Quark Vacuum Polarisation to Three Loops}\footnote{
     Work supported by BMFT under Contract 056KA93P6,
     DFG under Contract Ku502/6-1 and INTAS under Contract INTAS-93-0744.}
\end{LARGE}

\vspace{.8cm}

\begin{large}
 K.G.~Chetyrkin$^{a,b}$,
 J.H.~K\"uhn$^{b}$,
 M.~Steinhauser$^{b}$
\end{large}

\vspace{.5cm}

\begin{itemize}
\item[$^a$]
   Institute for Nuclear Research\\
   Russian Academy of Sciences, 60th October Anniversary Prospect 7a,\\
   Moscow 117312, Russia
\item[$^b$]
   Institut f\"ur Theoretische Teilchenphysik\\
   Universit\"at Karlsruhe, Kaiserstr. 12,    Postfach 6980,
   D-76128 Karlsruhe, Germany\\
\end{itemize}

\vspace{.5cm}

\begin{abstract}
\noindent
The real and imaginary part of the vacuum polarisation function
$\Pi(q^2)$ induced by a massive quark is calculated in perturbative
QCD up to order $\alpha_s^2$. The method is described and the
results are presented. This extends the calculation by
K\"all\'en and Sabry from two to three loops.
\end{abstract}


\vfill
\end{center}
\end{titlepage}

\section{Introduction}

The measurement of the total cross section for
electron positron annihilation into hadrons allows for
a unique test of pertubative QCD. The decay rate
$\Gamma(Z \to \mbox{hadrons})$ provides one of the
most precise determinations of the strong coupling
constant $\alpha_s$. In the high energy limit the quark masses
can often be neglected.
In this approximation QCD corrections to
$R  \equiv \sigma(e^+ e^- \to \mbox{hadrons})/
           \sigma(e^+ e^- \to \mu^+ \mu^-)$
have been calculated up to order $\alpha_s^3$
\cite{CheKatTka79DinSap79CelGon80,GorKatLar91SurSam91}.
For precision measurements the dominant mass corrections must be
included through an expansion in $m^2/s$. Terms up to order
$\alpha_s^3 m^2/s$
\cite{CheKue90}
and $\alpha_s^2 m^4/s^2$
\cite{CheKue94}
are available at present,
providing an acceptable approximation from the high energy
region down to intermediate energy values.
For a number of measurements, however, the information on the
complete mass dependence is desirable. This
includes charm and bottom meson production above the resonance
region, say
above $4.5$~GeV and $12$~GeV, respectively, and, of course,
top quark production at a future electron positron collider.

To order $\alpha_s$ this calculation was performed by
K\"all\'en and Sabry in the context of QED a long time ago
\cite{KaeSab55}.
With measurements of ever increasing precision, predictions
in order $\alpha_s^2$ are needed for a reliable
comparison between theory and experiment. Furthermore,
when one tries to apply the ${\cal O}(\alpha)$ result
to QCD, with its running coupling
constant, the choice of scale becomes important.
In fact, the distinction between the two intrinsically different
scales, the relative momentum versus the center of mass
energy, is crucial for a stable numerical prediction.
This information can be obtained from a full calculation
to order $\alpha_s^2$ only.
Such a calculation then allows to predict the cross section
in the complete energy region where perturbative QCD can be applied
--- from close to threshold up to high energies.

In this paper results for the cross section are presented in
order $\alpha_s^2$.
They are obtained from the vacuum polarisation $\Pi(q^2)$ which is
calculated up to three loops.
The imaginary part of the ``fermionic contribution'' ---  derived
from diagrams with a massless quark
loop inserted in the gluon propagator --- has been calculated in
\cite{HoaKueTeu95}.
All integrals could be performed to the end and the result was expressed
in terms of polylogarithms.
In this paper the calculation is extended to the full set of
diagrams relevant for QCD.
Instead of trying to perform the
integrals analytically, we
use the large $q^2$ behaviour of $\Pi(q^2)$ up to terms
of order $m^2/q^2$ and calculate
its Taylor series around $q^2=0$ up to terms of
order $q^8$. The leading and next-to-leading singularity is
deduced from the known behaviour of the nonrelativistic Green function
and the two-loop QCD potential.
Altogether eight constraints on $\Pi(q^2)$ are thus available,
four from $q^2=0$, two from $q^2\to-\infty$ and two from
the threshold.

The contributions $\sim C_F^2, \sim C_A C_F$ and $\sim C_F T n_l$
have to be treated separately since they differ significantly
in their  singularity structure.
For each of the three functions an interpolation is constructed
which incorporates all data and is based on conformal
mapping and Pad\'e approximation suggested in
\cite{FleTar94,BroFleTar93,BroBaiIly94,BaiBro95}.
Since the result for $C_F T n_l$ is available in closed form the
approximation method can be tested and shown to give excellent
result for this case. Reliable predictions for $R$ to order
$\alpha_s^2$ and arbitrary $m^2/s$ are thus available.

In this paper only results without renormalization group improvement
and resummation of the Coulomb singularities from higher orders
are presented. Resummation of leading higher order terms, phenomenological
applications and a more detailed discussion of our methods will be
presented elsewhere.

\section {The Calculation}

A large number of ingredients are needed in the
calculation of the three-loop vacuum polarisation function:
The high energy behaviour of $\Pi(q^2)$, its Taylor series at
$q^2=0$ and its singularities at threshold which are related to the QCD
potential.
The decomposition
of $\Pi$ according to its colour structure and the separation
of gluonic and fermionic contributions is crucial to display the
striking differences in the threshold region.
The approximation method is based on conformal mapping and
Pad\'e approximation.
In the present approach both
real and imaginary parts of the vacuum polarization
$\Pi(q^2)$ are calculated and its analyticity properties are exploited
heavily.

The physical observable $R(s)$ is related to $\Pi(q^2)$ by
\begin{eqnarray}
R(s)   &=&  12\pi\, \mbox{Im}\Pi(q^2=s+i\epsilon).
\end{eqnarray}
It is convenient to define
\begin{equation}
\Pi(q^2) = \Pi^{(0)}(q^2)
         + \frac{\alpha_s(\mu^2)}{\pi}\Pi^{(1)}(q^2)
         + \left(\frac{\alpha_s(\mu^2)}{\pi}\right)^2\Pi^{(2)}(q^2)
         + \cdots
\end{equation}
with the $\overline{\mbox{MS}}$ coupling $\alpha_s(\mu^2)$ defined in the
conventional way.

To describe the singularity structure of $\Pi$ in the region close to
threshold the perturbative QCD potential
\cite{Fis77}
\begin{eqnarray}
 V_{\mbox{\scriptsize QCD}}(\vec{q}\,^2) &=&
         -4\pi C_F\frac{\alpha_V(\vec{q}\,^2)}{\vec{q}\,^2},
\\
 \alpha_V(\vec{q}\,^2) &=&  \alpha_s(\mu^2)\Bigg[
      1 + \frac{\alpha_s(\mu^2)}{4\pi}\left(
          \left(\frac{11}{3}C_A-\frac{4}{3}T n_l\right)
          \left(-\ln\frac{\vec{q}\,^2}{\mu^2}+\frac{5}{3}\right)
          -\frac{8}{3}C_A            \right)
\label{alphav}
\Bigg]
\end{eqnarray}
will become important. The $C_A C_F$ and $C_F T n_l$
contributions have been displayed separately, with $C_F=4/3$,
$C_A=3$ and $T=1/2$.
The number of fermions is denoted by $N_f$ and has to be distinguished
from the number of light ($\equiv$ massless) fermions $n_l=N_f-1$.
To transform the results from QED to QCD the proper group
theoretical coefficients $C_F=1, C_A=0$ and $T=1$ have to be
used.

In this paper we are only concerned with contributions to
$\Pi(q^2)$ and $R(s)$ which originate from diagrams where
the electromagnetic current couples to the massive quark.
In order $\alpha_s$ and $\alpha_s^2$ all these amplitudes
are proportional to $Q_f^2$, the square of the charge
of the massive quark.
Diagrams where the electromagnetic current couples to a massless
quark and the massive quark is produced through a virtual
gluon have been calculated in
\cite{HoaKueTeu94} and will not be discussed here.

For the $\alpha_s^2$ calculation the following steps have
been performed:

\newpage

\vspace{.5em}
\noindent
{\bf Decomposition according to the colour structure}
\vspace{.5em}

The contributions from diagrams with $n_l$ light
or one massive
internal fermion loop will be denoted
by $C_F T n_l\Pi_{\mbox{\scriptsize\it l}}^{(2)}$ and
$C_F T\Pi_{\mbox{\scriptsize\it F}}^{(2)}$ with the group theoretical
coefficients
factored out. Purely gluonic corrections
are proportional to $C_F^2$ or $C_A C_F$. The former are the only
contributions in an abelian theory, the latter are characteristic for
the nonabelian aspects of QCD. It will be important in the
subsequent discussion to
treat these two classes separately, since they exhibit qualitatively
different behaviour close to threshold. The following decomposition
of $\Pi(q^2)$ (and similarly for $R(s)$)
is therefore adopted throughout the paper
\begin{eqnarray}
\Pi &=& \Pi^{(0)} + \frac{\alpha_s(\mu^2)}{\pi} \Pi^{(1)}
\\&&
       + \left(\frac{\alpha_s(\mu^2)}{\pi}\right)^2
         \left[
                C_F^2       \Pi_{\mbox{\scriptsize\it A}}^{(2)}
              + C_A C_F     \Pi_{\mbox{\scriptsize\it NA}}^{(2)}
              + C_F T n_l \Pi_{\mbox{\scriptsize\it l}}^{(2)}
              + C_F T     \Pi_{\mbox{\scriptsize\it F}}^{(2)}
         \right].
\end{eqnarray}
All steps described below have been performed seperately for
the first three contributions to $\Pi^{(2)}$.
In fact, new information is only
obtained for $\Pi_{\mbox{\scriptsize\it A}}^{(2)}$
and $\Pi_{\mbox{\scriptsize\it NA}}^{(2)}$ since
$\mbox{Im}\Pi_{\mbox{\scriptsize\it l}}^{(2)}$
is known analytically already.
The amplitude with a massive internal fermion exhibits a two
particle cut with threshold at $2m$ which has been calculated analytically
\cite{HoaKueTeu95}.
The contribution from a four particle cut
with threshold at $4m$ is given
in terms of a two dimensional integral
\cite{HoaKueTeu95}
which can be solved
easily numerically.
The $\Pi_F^{(2)}$ term will not be treated in this paper.

\vspace{.5em}
\noindent
{\bf Analysis of the high $q^2$ behaviour}
\vspace{.5em}

The high energy behaviour of $\Pi$ provides important
constraints on the complete answer.
In the limit of small $m^2/q^2$ the constant term and the one
proportional to $m^2/q^2$ (modulated by powers of $\ln \mu^2/q^2$) have been
calculated a long time ago
\cite{GorKatLar86}.
For the imaginary part even the $m^4/q^4$ terms are available
\cite{CheKue94}.
This provides an important test of the numerical results
presented below.

\vspace{.5em}
\noindent
{\bf Threshold behaviour}
\vspace{.5em}

General arguments based on the influence of Coulomb exchange close to
threshold, combined with the information on the perturbative QCD
potential and the running of $\alpha_s$ dictate the singularities
and the structure of the leading cuts close to threshold, that
is for small $v=\sqrt{1-4m^2/s}$.
The $C_F^2$ term
is directly related to the QED result with internal photon lines only.
The leading $1/v$
singularity  and the constant term  of $R_A$
can be predicted from the nonrelativistic
Greens function for the Coulomb potential
and the ${\cal O}(\alpha_s)$ calculation.
The next-to-leading
term is determined by the combination of one loop
results again with the Coulomb singularities
\cite{BarGatKoeKun75,VolSmi94,BaiBro95}. One finds
\begin{eqnarray}
R_{\mbox{\scriptsize\it A}}^{(2)} &=&
3\left(\frac{\pi^4}{8v} - 3\pi^2 + \ldots\right).
\label{Ra}
\end{eqnarray}

The contributions $\sim C_A C_F$ and $\sim C_F T n_l$ can be treated
in parallel.
The leading $C_AC_F$ and $C_F T n_l$ term in $R$ is proportional
to $\ln v^2$ and is responsible for the evolution of the
coupling constant close to threshold. Also the constant term can
be predicted by the observation, that the leading term in
order $\alpha_s$ is induced by the potential.
The ${\cal O}(\alpha_s)$ result
\begin{eqnarray}
R&=&3\left(\frac{3}{2}v+ C_F\frac{3\pi^2}{4}\frac{\alpha_s}{\pi}+\ldots
     \right)
\end{eqnarray}
is employed to predict the logarithmic and constant $C_FC_A$ and
$C_FTn_l$ terms of ${\cal O}(\alpha_s^2)$ by replacing $\alpha_s$
by $\alpha_V(4\vec{p}\,^2=v^2 s)$ as given in Eq.(\ref{alphav}).
This implies the following threshold behaviour:
\begin{eqnarray}
R_{\mbox{\scriptsize\it NA}}^{(2)}&=&3\pi^2\left(\frac{31}{48}
                                            -\frac{11}{16}\ln v^2
                                            +\frac{11}{16}\ln\frac{\mu^2}{s}
                                   +\ldots    \right),
\label{Rna}\\
R_{\mbox{\scriptsize\it l}}^{(2)}&=&3\pi^2\left(-\frac{5}{12}
                                       +\frac{1}{4}\ln v^2
                                       +\frac{1}{4}\ln\frac{\mu^2}{s}
                                   +\ldots    \right).
\label{Rnl}
\end{eqnarray}

This ansatz can be verified for the $C_F T n_l$ term since in this case
the result is known in analytical form
\cite{HoaKueTeu95}.
Extending the ansatz from the behaviour of the imaginary part
close to the branching point into the complex plane
allows to predict the leading term of
$\Pi(q^2)$ $\sim \ln v$ and $\sim \ln v^2$.

\vspace{.5em}
\noindent
{\bf Behaviour at $q^2=0$}
\vspace{.5em}

Important information is contained in the Taylor series of $\Pi(q^2)$
around zero. The calculation of the first four nontrivial terms is
based on the evaluation of three-loop tadpole integrals with
the help of the algebraic program MATAD written in FORM
\cite{VerFORM}
which
performs the traces,
calculates the derivatives with respect to the external momenta.
It reduces the large number of different
integrals to one master integral and a few simple ones
through an elaborate set of
recursion relations based on the integration-by-parts method
\cite{CheTka81,Bro92}.
Though this master integral, calculated in
\cite{Bro92}, appears in the result for single
diagrams, in the final expession it cancels.
The following results are obtained
(excluding the $\Pi_F^{(2)}$ term):
\begin{eqnarray}
\Pi^{(2)} &=&
              \frac{3}{16\pi^2}
              \sum_{n>0} C_{n} \left(\frac{q^2}{4m^2}\right)^n,
\\
C_{1} &=& C_F^2\left(
  -\frac{8687}{864}   + 4\zeta(2) - \frac{32}{5}\zeta(2)\ln2
  + \frac{22781}{1728}\zeta(3)
               \right)
\nonumber\\
&+& C_A C_F\left(
     \frac{127}{192} + \frac{902}{243}\lmm
   - \frac{16}{15}\zeta(2) + \frac{16}{5}\zeta(2)\ln2
   + \frac{1451}{384}\zeta(3)
         \right)
\nonumber\\
&+& C_F T n_l\left(
   - \frac{142}{243} - \frac{328}{243}\lmm - \frac{16}{15}\zeta(2)
             \right),
\nonumber\\
C_{2} &=& C_F^2\left(
  - \frac{223404289}{1866240}
  + \frac{24}{7}\zeta(2) - \frac{192}{35}\zeta(2)\ln2
  + \frac{4857587}{46080}\zeta(3)
               \right)
\nonumber\\
&+& C_A C_F\left(
  - \frac{1030213543}{93312000} + \frac{4939}{2025}\lmm
  - \frac{32}{35}\zeta(2) + \frac{96}{35}\zeta(2)\ln2
  + \frac{723515}{55296}\zeta(3)
         \right)
\nonumber\\
&+& C_F T n_l\left(
  - \frac{40703}{60750} - \frac{1796}{2025}\lmm
  - \frac{32}{35}\zeta(2)
             \right),
\nonumber\\
C_{3} &=& C_F^2\left(
  - \frac{885937890461}{1161216000}
  + \frac{64}{21}\zeta(2)
  - \frac{512}{105}\zeta(2)\ln2
  + \frac{33067024499}{51609600}\zeta(3)
               \right)
\nonumber\\
&+& C_A C_F\left(
  - \frac{95905830011197}{1706987520000}
  + \frac{2749076}{1488375}\lmm
  - \frac{256}{315}\zeta(2)
  + \frac{256}{105}\zeta(2)\ln2
         \right.
\nonumber\\
  &+& \left. \frac{5164056461}{103219200}\zeta(3)
         \right)
+ C_F T n_l\left(
  - \frac{9703588}{17364375} - \frac{999664}{1488375}\lmm
  - \frac{256}{315}\zeta(2)
             \right),
\nonumber\\
C_{4} &=& C_F^2 \left(
  - \frac{269240669884818833}{61451550720000}
  + \frac{640}{231}\zeta(2) - \frac{1024}{231}\zeta(2)\ln2
  + \frac{1507351507033}{412876800}\zeta(3)
                \right)
\nonumber\\
&+& C_A C_F\left(
   - \frac{36675392331131681}{158018273280000}
   + \frac{571846}{382725}\lmm
   - \frac{512}{693}\zeta(2) + \frac{512}{231}\zeta(2)\ln2
         \right.
\nonumber\\
   &+& \left.
       \frac{1455887207647}{7431782400}\zeta(3)
         \right)
+ C_F T n_l\left(
  - \frac{54924808}{120558375} - \frac{207944}{382725}\lmm
  - \frac{512}{693}\zeta(2)
             \right).
\nonumber
\end{eqnarray}
The $C_F^2$ terms in $C_1$, $C_2$ and $C_3$ are in agreement with
\cite{BaiBro95}, the remaining ones are new.
(Note that the $C_F T$ contribution calculated in
\cite{BaiBro95} were obtained with massive internal fermions.)

\vspace{.5em}
\noindent
{\bf Conformal mapping and Pad\'e approximation}
\vspace{.5em}

The vacuum polarisation function
$\Pi^{(2)}$ is analytic in the complex plane
cut from $q^2=4m^2$ to $+\infty$. The Taylor series in $q^2$ is
thus convergent in the domain $|q^2|<4m^2$ only. The
conformal mapping
which corresponds to the variable
transformation
\begin{eqnarray}
\omega = \frac{1-\sqrt{1-q^2/4m^2}}{1+\sqrt{1-q^2/4m^2}},\,\,\,\,
&&
\frac{q^2}{4m^2} = \frac{4\omega}{(1+\omega)^2},
\label{omega}
\end{eqnarray}
transforms the cut complex $q^2$ plane into the
interior of the
unit circle. The special points
$q^2=0,4m^2,-\infty$ correspond to $\omega=0,1,-1$, respectively.

The upper (lower) part of the cut is mapped onto the upper (lower)
perimeter of the circle.
The Taylor series in $\omega$ thus converges in the interior of the
unit circle. To obtain predictions for
$\Pi(q^2)$ at the boundary it has been suggested
\cite{FleTar94,BroFleTar93}
to use
the Pad\'e approximation
which converges towards $\Pi(q^2)$ even on the perimeter.
To improve the accuracy
the singular threshold behaviour
and the large $q^2$ behaviour
is incorporated into simple analytical functions
which are removed
from $\Pi^{(2)}$ before the Pad\'e approximation is
performed.
The quality of this
procedure can be tested by comparing the prediction with
the known result for $\mbox{Im}\Pi_{\mbox{\scriptsize\it l}}^{(2)}$.

The logarithmic singularities at threshold and large $q^2$
are removed by subtraction, the $1/v$ singularity, which is present
for the $C_F^2$ terms only, by multiplication with $v$ as
suggested in
\cite{BaiBro95}.
The imaginary part of the remainder which is actually
approximated by the Pad\'e method is thus smooth in the
entire circle, numerically small and vanishes at
$\omega=1$ and $\omega=-1$.

\section{Results}

After performing the Pad\'e approximation for the smooth remainder
with $\omega$ as natural variable, the transformation (\ref{omega})
is inverted and the full vacuum polarisation function reconstructed
by reintroducing the threshold and high energy terms. This
procedure provides real and imaginary parts of $\Pi^{(2)}$.
Subsequently only the absorbtive part of $\Pi^{(2)}$ (multiplied by
$12\pi$) will be presented.

\begin{figure}[ht]
 \begin{center}
 \begin{tabular}{c}
   \epsfxsize=11.5cm
   \leavevmode
   \epsffile[110 330 460 520]{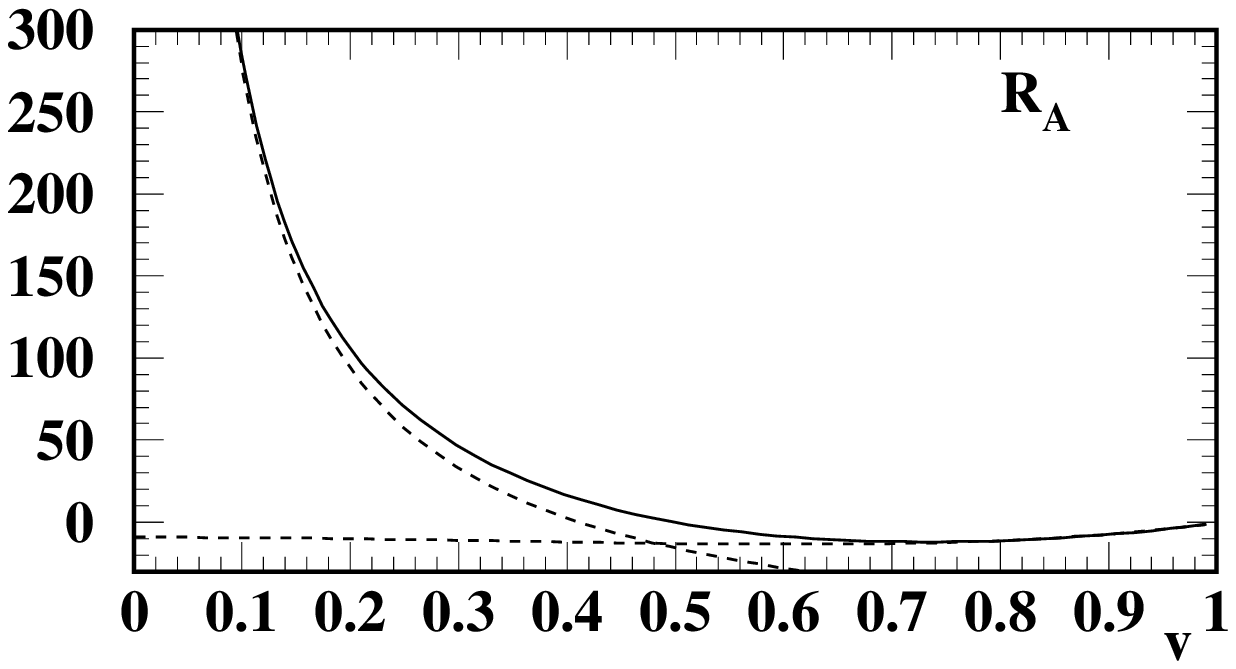}
   \\
   \epsfxsize=11.5cm
   \leavevmode
   \epsffile[110 330 460 520]{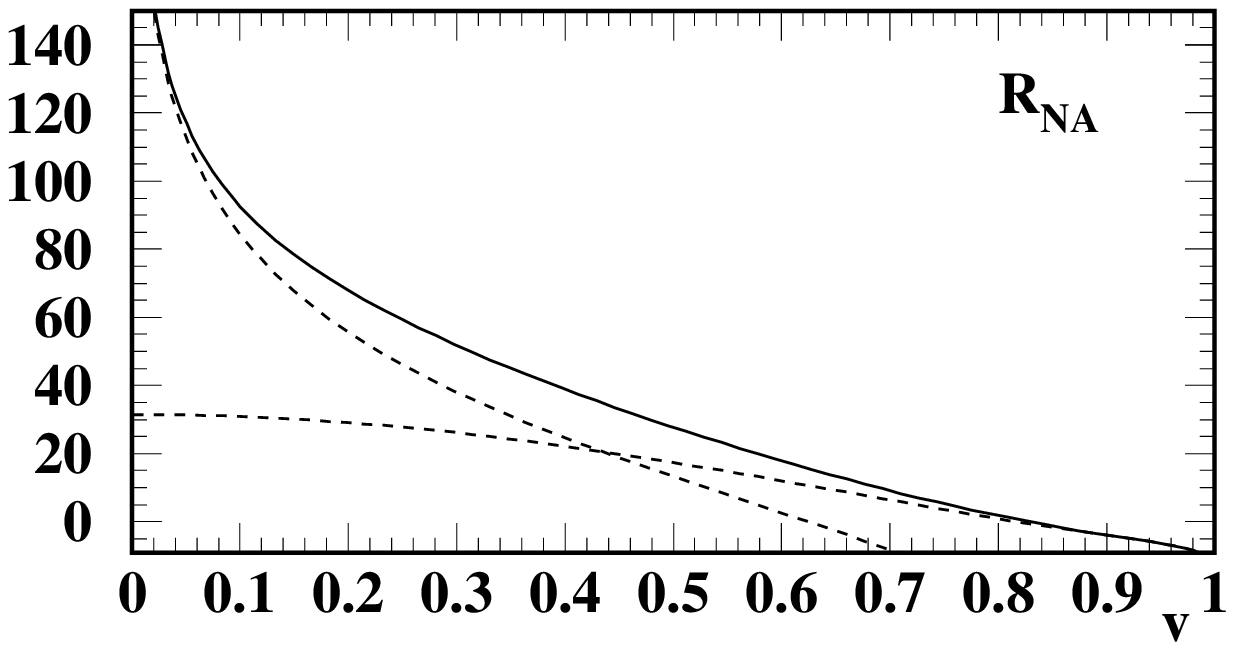}
   \\
   \epsfxsize=11.5cm
   \leavevmode
   \epsffile[110 330 460 520]{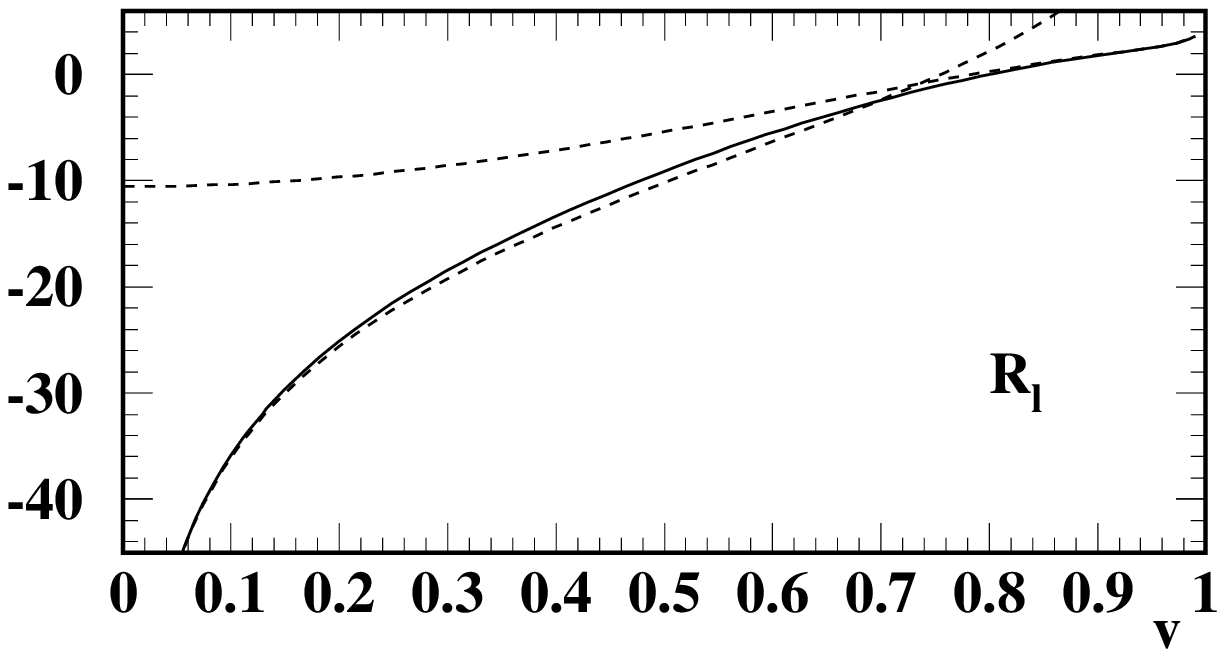}
 \end{tabular}
 \caption{\label{figfull} Complete results
                          plotted against
                          $v=\protect\sqrt{1-4m^2/s}$. The
                          high energy approximation includes the $m^4/s^2$
                          term.}
 \end{center}
\end{figure}

\begin{figure}[ht]
 \begin{center}
 \begin{tabular}{c}
   \epsfxsize=11.5cm
   \leavevmode
   \epsffile[110 330 460 520]{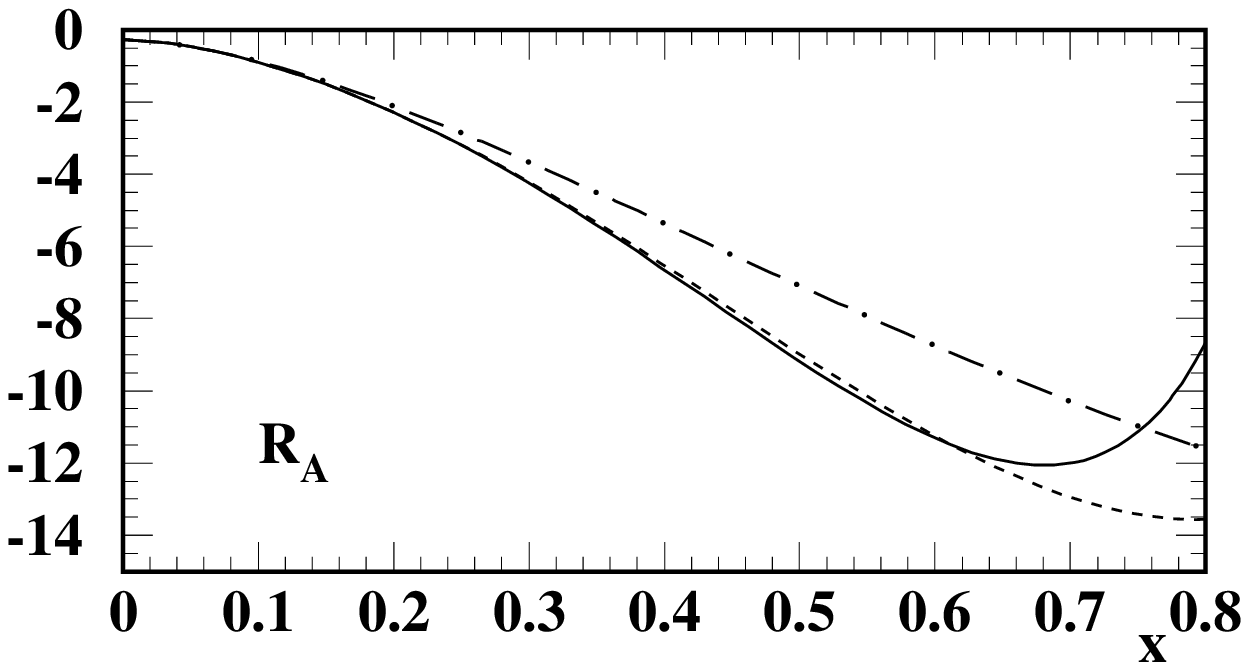}
   \\
   \epsfxsize=11.5cm
   \leavevmode
   \epsffile[110 330 460 520]{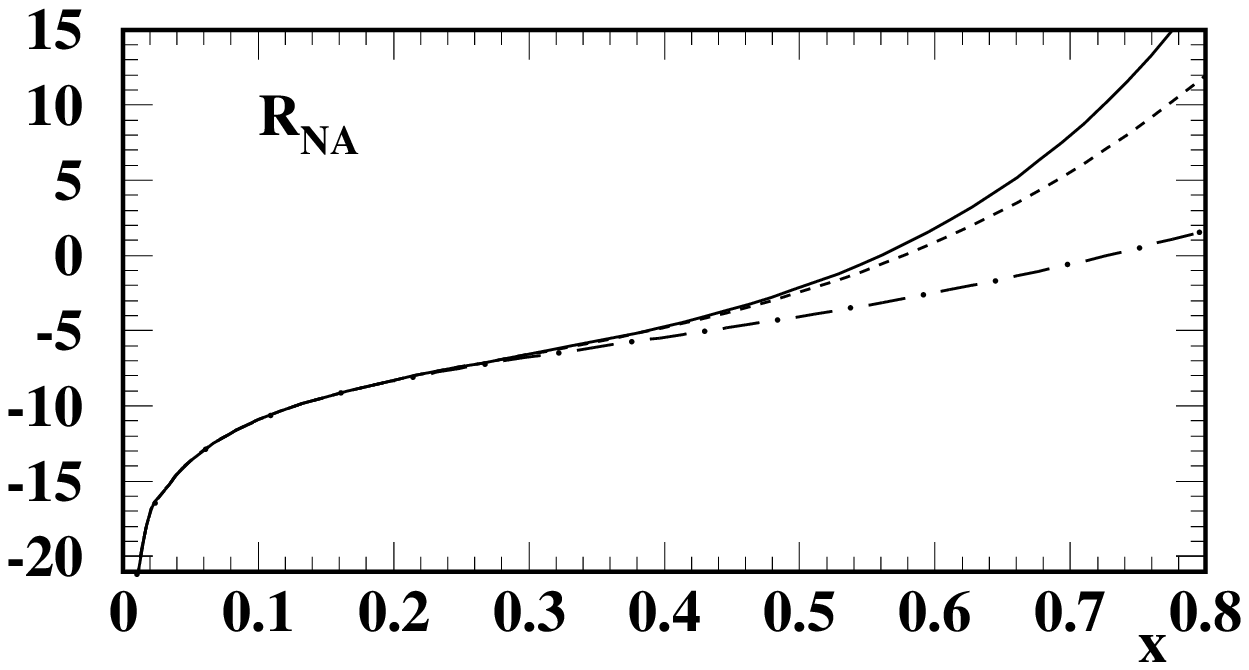}
   \\
   \epsfxsize=11.5cm
   \leavevmode
   \epsffile[110 330 460 520]{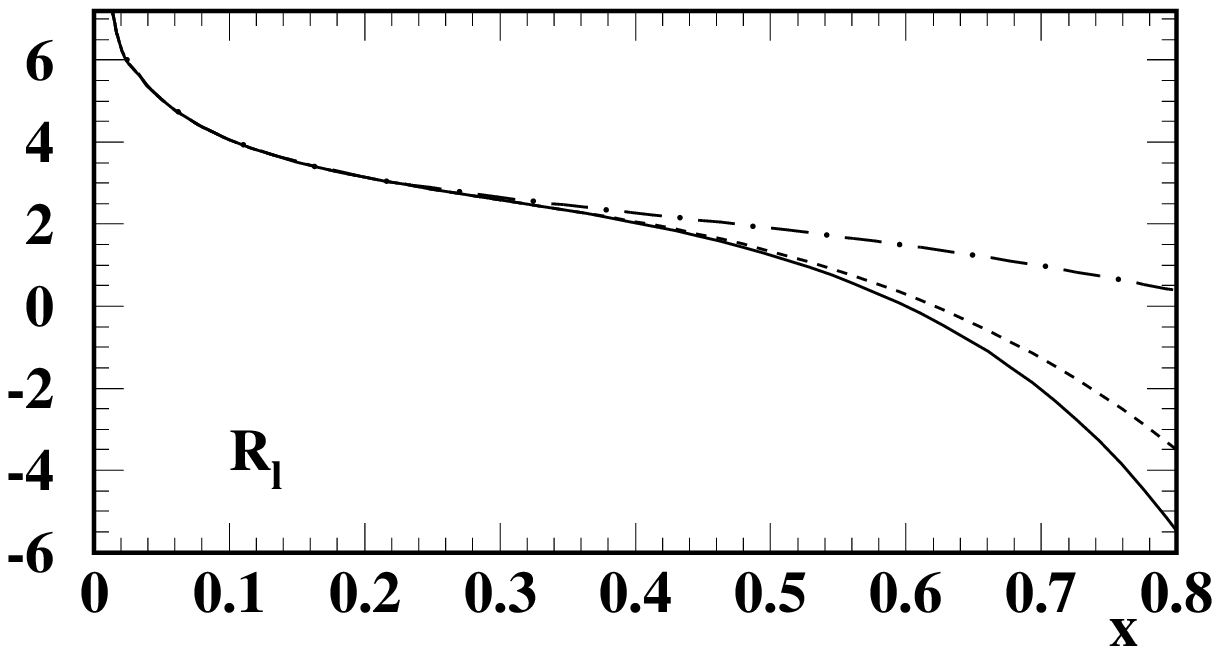}
 \end{tabular}
 \caption{\label{fighigh}High energy region.
                         The complete results (full line) are
                         compared to the high energy
                         approximations including the $m^2/s$
                         (dash-dotted) and the $m^4/s^2$ (dashed) terms.}
 \end{center}
\end{figure}

In Figure \ref{figfull} the complete results are shown
for $\mu^2=m^2$ with
$R_{\mbox{\scriptsize\it A}}^{(2)}$, $R_{\mbox{\scriptsize\it NA}}^{(2)}$
and $R_{\mbox{\scriptsize\it l}}^{(2)}$
displayed separately. The solid curves are based on the
$[4/2]$, $[3/2]$ and $[3/2]$ Pad\'e approximants for $A$, $NA$ and $l$,
respectively. The threshold and high energy behaviour
is given by the dashed curves. The exact analytical
result which is known for the $R_{\mbox{\scriptsize\it l}}^{(2)}$ contribution
only differs from the approximation curve in Figure \ref{figfull}
by less than the thickness of the line. The quality of the
approximation for $R_{\mbox{\scriptsize\it A}}^{(2)}$ and
$R_{\mbox{\scriptsize\it NA}}^{(2)}$
is confirmed by the comparison of the high energy behaviour
of the approximation with the known asymptotic behaviour
(Figure \ref{fighigh}). The quadratic approximation
(dash-dotted line) is incorporated into $R^{(2)}$ by construction,
the quartic approximation shown (dashed line) is known from
\cite{CheKue94}, but is evidently very well recovered by the
method presented here.

\begin{figure}[ht]
 \begin{center}
 \begin{tabular}{c}
   \epsfxsize=11.5cm
   \leavevmode
   \epsffile[110 330 460 520]{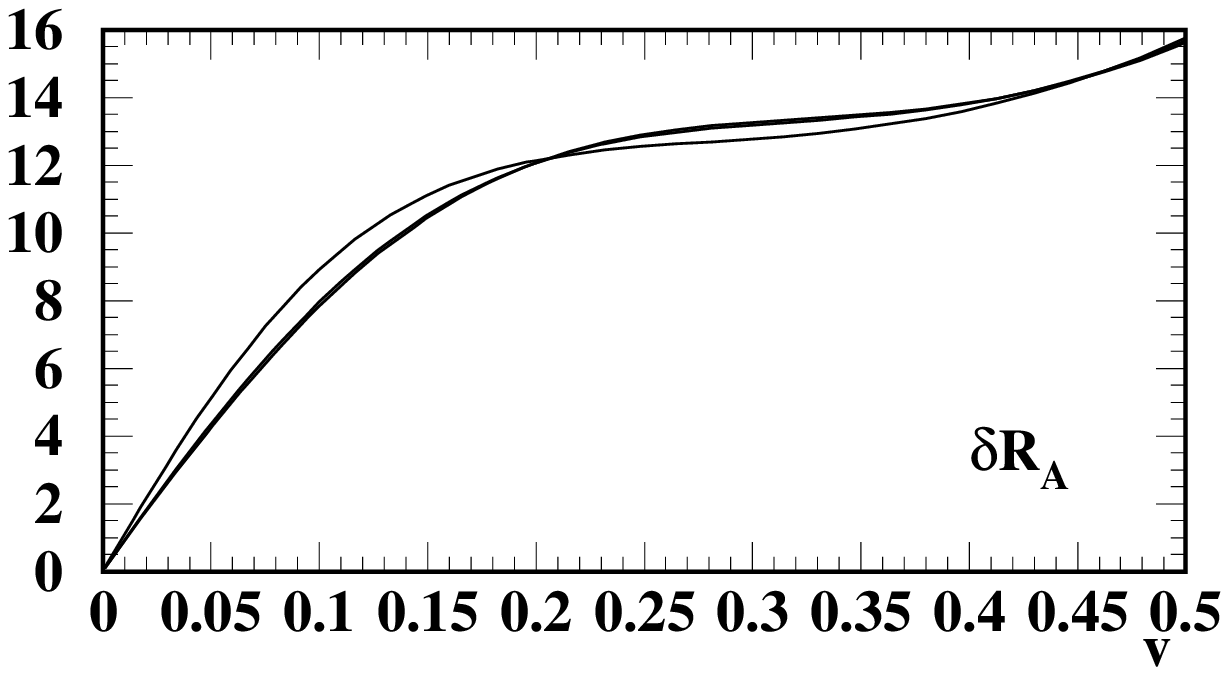}
   \\
   \epsfxsize=11.5cm
   \leavevmode
   \epsffile[110 330 460 520]{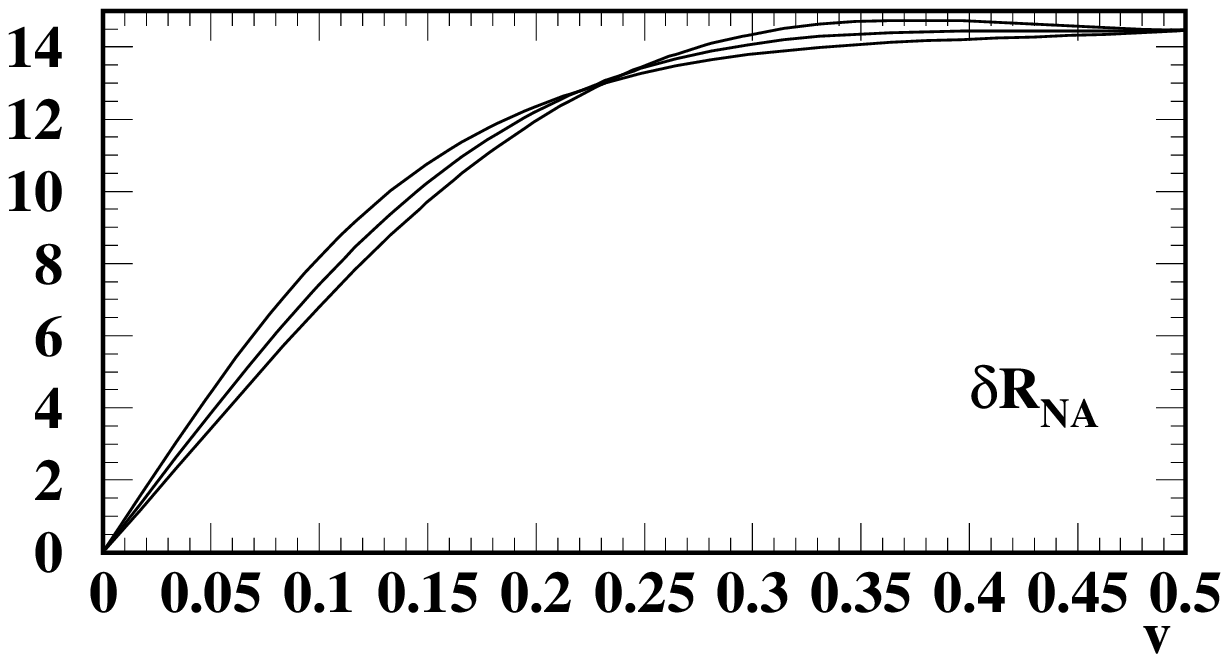}
   \\
   \epsfxsize=11.5cm
   \leavevmode
   \epsffile[110 330 460 520]{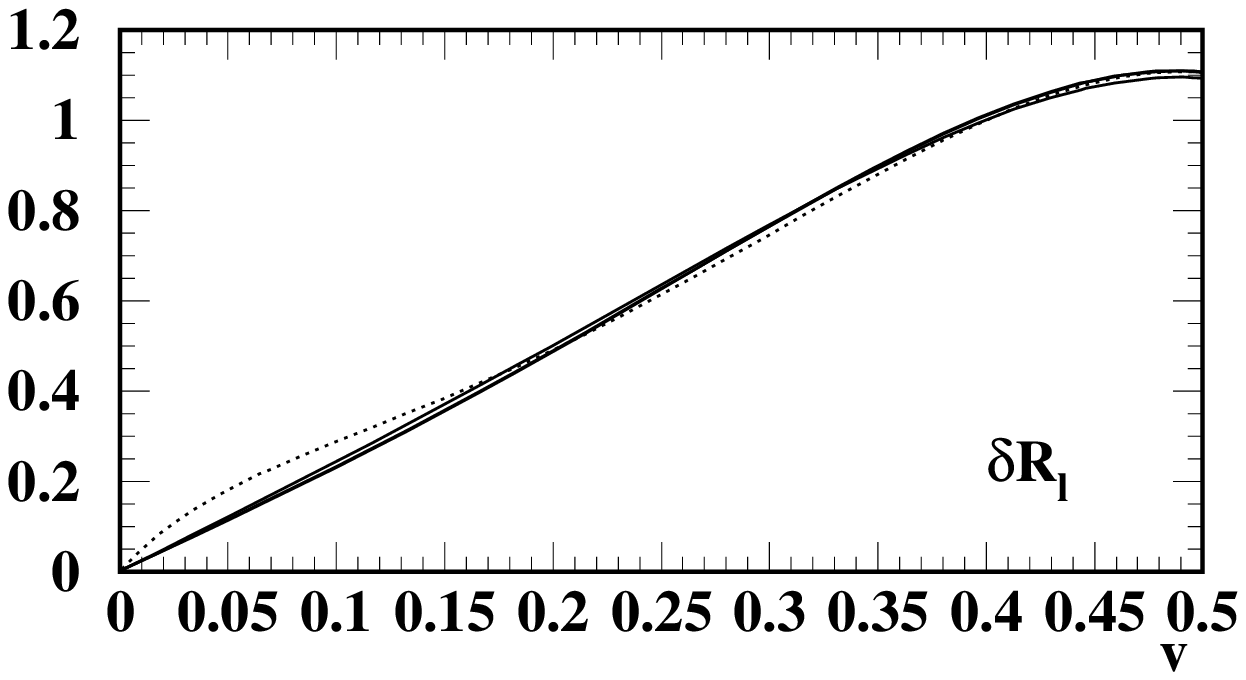}
 \end{tabular}
 \caption{\label{figthr}Threshold behaviour of the remainder $\delta R$
                        for three different Pad\'e approximants. (The singular
                        and constant parts around threshold are subtracted.)}
 \end{center}
\end{figure}

Different Pad\'e approximations of the same degree
($[4/2]$ or $[2/4]$) and approximants with a
reduced number of parameters give rise to practically indentical
predictions, which could hardly be distinguished in
Figures \ref{figfull} and \ref{fighigh}. Minor variations are observed close
to threshold, {\it after} subtracting the singular and constant parts.
The remainder $\delta R$ for three different Pad\'e
approximants is shown in
Figure \ref{figthr}. The stability of the
different approximations and their smooth behaviour close to the point
$v=1$ can be considered as additional confirmation of our ansatz
in Eqs.(\ref{Ra},\ref{Rna},\ref{Rnl}).
The spread of the different curves can be used to estimate
the quality of the approximation.
For $R_{\mbox{\scriptsize\it l}}^{(2)}$ the approximation can be
compared to the exact result (dotted line) and the
nearly perfect agreement
gives
additional support to the approach presented in this work.

To summarize: Real and imaginary part of the vacuum polarisation
function $\Pi(q^2)$ from a massive quark have been calculated up to
three loops for QCD and QED. This result extends the classic
calculations  of K\"allen and Sabri \cite{KaeSab55} to
next-to-leading order. The imaginary part
can be used to predict the cross section for production of
massive quarks for arbitrary $m^2/s$, wherever perturbative QCD can be
justified --- from above the quarkonium resonance region up to high
energies.

\vspace{5ex}
{\bf Acknowledgments}

\noindent
We would like to thank A.H. Hoang and T. Teubner for
many interesting discussions. Their programs with the
analytical results for
$R_{\mbox{\scriptsize\it l}}^{(2)}$ were crucial for our tests
of the approximation methods.
We are grateful to D. Broadhurst for helpful discussions.
J.H.K. would like to thank
S. Brodsky for discussions of the V-scheme and the threshold
behaviour and K.G.Ch. thanks A.A. Pivovarov for careful
reading the manuscript.


\newpage

\begin{thebibliography}{99}

\bibitem{CheKatTka79DinSap79CelGon80}
K.G. Chetyrkin, A.L. Kataev, F.V. Tkachov,  Phys.\ Lett.
\ {\bf B85}  (1979) 277;\\
M. Dine, J. Sapirstein, Phys.\ Rev.\ Lett. {\bf 43} (1979) 668; \\
W. Celmaster, R.J. Gonsalves, Phys.\ Rev.\ Lett. {\bf 44} (1980) 560.
\bibitem{GorKatLar91SurSam91}
S.G. Gorishny, A.L. Kataev, S.A. Larin, Phys.\ Lett. {\bf B259} (1991) 144;\\
L.R. Surguladze, M.A. Samuel,  Phys. Rev. Lett. {\bf 66} (1991) 560;
erratum ibid, 2416.

\bibitem{CheKue90}
K.G. Chetyrkin, J.H. K\"uhn, Phys.\ Lett.\ {\bf B248} (1990) 359.

\bibitem{CheKue94}
K.G. Chetyrkin, J.H. K\"uhn, Nucl.\ Phys.\ {\bf B432} (1994) 337.

\bibitem{KaeSab55}
G. K\"allen and A. Sabry, K. Dan. Vidensk. Selsk. Mat.-Fys. Medd.
{\bf 29} (1995) No. 17,\\
see also
J. Schwinger, Particles, {\it Sources and Fields}, Vol.II,
   (Addison-Wesley, New York, 1973).

\bibitem{HoaKueTeu95}
A.H. Hoang, J.H. K\"uhn, T. Teubner, Nucl.Phys. {\bf B452} (1995) 173.

\bibitem{FleTar94}
J. Fleischer and O.V. Tarasov, Z. Phys. {\bf C64} (1994) 413.

\bibitem{BroFleTar93}
D.J. Broadhurst, J. Fleischer and O.V. Tarasov, Z. Phys. {\bf C60}
(1993) 287.

\bibitem{BroBaiIly94}
D.J. Broadhurst et. al., Phys. Lett. {\bf B329} (1994) 103.

\bibitem{BaiBro95}
P.A. Baikov and D.J. Broadhurst, Report Nos. OUT-4102-54, INP-95-13/377 and
hep--ph/9504398, to appear in New Computing Techniques in Physics
Research IV, (World Scientific, in press).

\bibitem{Fis77}
W. Fischler, Nucl. Phys. {\bf B129} (1977) 157.\\
A. Billoire, Phys. Lett. {\bf B92} (1980) 343.

\bibitem{HoaKueTeu94}
A.H. Hoang, M. Je\.zabek, J.H. K\"uhn, T. Teubner,
Phys.\ Lett.\ {\bf B338} (1994) 330.

\bibitem{GorKatLar86}
S.G. Gorishny, A.L. Kataev and S.A. Larin,
Nuovo Cimento {\bf 92 A} (1986) 119.

\bibitem{BarGatKoeKun75}
R. Barbieri, R. Gatto, R. K\"ogerler and Z. Kunszt, Phys.\ Lett.\
{\bf B57} (1975) 455.

\bibitem{VolSmi94}
B. H. Smith and M.B. Voloshin,
Phys.\ Lett.\ {\bf B324} (1994) 117.

\bibitem{VerFORM}
J.A.M. Vermaseren, {\it Symbolic Manipulation with FORM},
(Computer Algebra Netherlands, Amsterdam, 1991).

\bibitem{CheTka81}
F.V. Tkachov, Phys.\ Lett.\ {\bf B100} (1981) 65.\\
K.G. Chetyrkin and F.V. Tkachov, Nucl. Phys. {\bf B192} (1981) 159.

\bibitem{Bro92}
D.J. Broadhurst, Z. Phys. {\bf C54} (1992) 54.

\end{thebibliography}
\end{document}